\documentstyle[aas2pp4]{article}

\slugcomment{Submitted to ApJSS.}

\lefthead{de Mello et al.}
\righthead{Faint Interacting Galaxies}

\begin{document}

\title{A Catalog of Faint Interacting Galaxies in Pairs and Groups
\footnote{The catalogs can be obtained by email: duilia@on.br}}

\author{Du\'\i lia F. de Mello\altaffilmark{2,3,4}} 

\affil{CTIO, La Serena, Chile and ON, RJ, Brazil}

\author{Leopoldo Infante}

\affil{P. Universidad Cat\'olica de Chile}

\and 

\author{Felipe Menanteau}

\affil{P. Universidad Cat\'olica de Chile}

\altaffiltext{2}{FAPESP Fellow during 1995} 
\altaffiltext{3}{Current CNPq Fellow}
\altaffiltext{4}{Present address: Observat\'orio Nacional, Rio, 20921-400 RJ Brazil}

\begin{abstract}

We have carried out an extensive survey of faint galaxies in order to examine
the rise in  the merger rate with redshift and to study the statistical
relations between  close interacting galaxies and the field galaxy population.
In this paper  we present the catalogs of faint pairs and groups of galaxies of
46 equatorial fields taken with the CTIO 4m prime focus. The data set contains 73,988 galaxies covering a total area of
2.23~deg$^{2}$. We have found 1751 isolated pairs and 30 groups
of galaxies within  $19<m_{R}<22$ and $2''<\theta<6''$ in this area. Our results 
show clearly an increase in pairs and groups of galaxies in comparison to a randomly 
generated catalog.

\end{abstract}

\keywords{galaxies: interactions --- galaxies: evolution --- catalogs}

\bigskip

\bigskip

\bigskip 

{\centerline {\it Astrophysical Journal Supplement Series: Accepted}}

\section{Introduction}

Observations of pairs of galaxies at intermediate redshifts have revealed a
larger number of objects in the past (Zepf \& Koo 1989; Burkey et al. 1994;
Carlberg et al. 1994 ) suggesting that merging has possibly an important role in
galaxy evolution. This is verified, for example, by the excess of blue
star-forming galaxies at intermediate redshifts (Butcher \& Oemler 1984).
However, most of the previous works are limited to small samples of objects
covering a small area in the sky and making their statistics not ideal.  
There are a few samples of objectively selected nearby pairs and groups of
galaxies (Karachentsev 1972; Hickson 1982; Maia et al. 1994; Prandoni
et al. 1994; Soares et al. 1995; Reduzzi \& Rampazzo 1995). On the other hand, intermediate redshifts lacks of such
extensive survey since this type of survey requires large telescopes 
equipped with modern CCD detectors. Therefore, the number of faint galaxies with 
known redshift is still low (see Koo \& Kron 1992 and Ellis 1995 for reviews on redshift surveys and imaging of the faint galaxy population; Carlberg et al. 1994; Yee \& Ellingson 1995 ). 
We have started a long term project in order to improve this situation. 
In this paper, we present an extensive catalog of
faint interacting galaxies in pairs and groups. We believe that catalogs like these are one of the first steps towards understanding galaxy evolution. A following step will be to measure redshifts for a statistically significant number of galaxies which will require a large amount of observing time at 4-m class telescopes. However, our sample is equatorial and accessible from both hemispheres. 

\section{Observations and Reductions}

The observations were taken at the CTIO 4m prime focus camera by the  High-Z
Supernovae Search Group  with a redshifted filter B
which is almost equivalent to a regular Kron-Cousin  R filter (B/($z=0.4$)). The
image set comprises of 46 equatorial images  of 15$'\times 15'$ (0.44$''$
pixel$^{-1}$) making a total area of 2.8532 deg$^{2}$. Single 5 minutes exposures
were sufficient to provide good quality images. 

To convert instrumental magnitude into  m$_{R}$ we used the equation m$_{R}$= 31.5 -
2.5log(counts) - 0.1X, where X is the airmass. The photometric zero point was
calibrated by observing a m$_{R}$=16.97 star at R.A.=10$^{h}$ 50$^{m}$ 49$^{s}$ and DEC.= -9$^{\circ} 14' 31.4''$ (epoch 2000), with the AAT and the MSSSO
2.3m on two photometric nights (Schmidt 1995).  

The CCD images were first convolved with a lowered Gaussian kernel and then
images were detected by using a local maximum technique.  The arrays were
background subtracted to have zero mean (i.e., local sky has already been
subtracted), and smoothed or cross-correlated. There are two conditions which
must be satisfied, the intensity of the pixel must be above a predetermined
threshold ($1.5\sigma$) and that the pixel must be a local maximum.  The
algorithm then combines all local maxima within N contiguous pixels of each
other and  compute centroids of each final local maximum. Finally, objects which
are closer to each other than a given tolerance are merged. Masks were made to
exclude zones with bright objects and bad pixels. A total area of
0.2372 deg$^{2}$ was excluded. 

The photometry was performed by an algorithm which measures the ``total'' light
within a variable aperture (Kron 1980; Infante 1987) which is better for
extended objects than fixed circular apertures. Objects were then classified as
galaxies, stars or noise by using the properties of the inverse first and second
moments of the images which gives a measure of intrinsic size and central
compactness, respectively. 

\section{Completeness}

Simulations were performed in order to test detection and photometry of faint
images as a function of magnitude for different types of data - that is to say,
different observing conditions and object type (i.e., stars, disk galaxies
and spheroidal galaxies). These simulations were carried out in the same way as
in Infante (1987). Stars and galaxies were simulated on top of flat noisy frames
and then detection and photometry routines were run. The tests were designed so
that the simulations resemble as much as possible the data described above. (For more
details see Infante 1987).

The noise frames ($500\time500$ pixels) were created to have the same
characteristics as the real frames. These were as follows:   
pixel size = $0.44''$, sky background = 4100 DU, noise = 40 DU/pixel, FWHM = $2''$,
read-out-noise = 4.2 [e], gain = 2.9 [e/DU]. The pixel values of the noise
frames were drawn from a gaussian distribution; the noise being the dispersion
and the sky background being the mean.  Stars, disk galaxies and spheroidal
galaxies were simulated in turn.  50 paired objects of the same class, with the
same parameters, were created at random positions on the noise frames. The
detection and photometry algorithms were then run. The variables were the
magnitude ($20 \leq m_R \leq 24$), the seeing ($1'' \leq seeing \leq 1.6''$) and
the pair separation ($1'' \leq \theta \leq 6''$). The redshift determines the
angular size of the galaxies through the relativistic angular size-distance
relation.

\subsection{Limiting Magnitude }

The detection results for stars and spheroidal galaxies are shown in 
Fig. \ref{fig:comp-mag}. In all cases, this magnitude is a function of sky surface
brightness. However, for our data it spans less than 0.2 magnitudes in $m_R$.  For
unresolved objects (FWHM$<~1.3''$) the completeness limit is $m_R\approx 23.0$ at
$\mu_{m_R}\approx 20.7$, which is the case for the bulk of our observations. The
limiting magnitude for resolved objects depends on the surface brightness of the
object. For a fixed magnitude the surface brightness goes down with diameter,
hence distance. Thus, the limiting magnitude for galaxies is a function of
redshift.  For both, disk and spheroidal galaxies, at $z\leq0.3$, the 99\%
completeness limit  is $m_R\approx22.8$ for $\mu_{m_R}\approx20.7$, very close 
to what we obtained for stars.  

\placefigure{\ref{fig:comp-mag}}
 
The turn over of galaxy number counts as a function of magnitude provides a
rough estimate of completeness as well. It is well known that galaxy number
counts in $R$ rises as dlogN/dm~$\approx 0.3$ up to $R~<~25$ (Infante et al.
1986). Any turn over in the number counts at $R < 25$ must be due to
incompleteness.  Galaxy number counts as a function of $m_R$ are shown in Fig.
\ref{fig:counts}. The slope of the counts is  d(logN)/dm = 0.33 in agreement
with Infante et al. (1986). It is clear from this figure that the turn over in
the number counts occurs at $m_R>22.5$. We, therefore, claim that our catalog
of pairs and groups is 99\% complete at $m_R<22$.

\placefigure{\ref{fig:counts}}

\subsection{Separation}

We now turn to determine the limit at which our finding algorithm is able to
resolve galaxies separated by a given angular distance. For each simulation, on
a 500 $\times $ 500 noise frame,  50 pairs of objects (resolved and unresolved)
were placed at random positions. The variables were the seeing and the angular
separation. Simulations were carried out for $m_R$=20 and $m_R$=22 in order to
detect any brightness dependence.  After 50 runs we conclude that both objects
in pairs separated by more than $2''$ are always detected for $1.0''\leq seeing
\leq 1.6''$. This limit does not change significantly with brightness,
for unresolved (stars) and for resolved (spheroidal galaxies) objects.

\section{Selection Criteria}

Our selection criteria are based on separations between pairs of galaxies. Pairs
of faint galaxies (separation $\Delta\theta$) are chosen such that
$\theta_{min}~<~\Delta\theta~<~\theta_{max}$.  Here $\theta_{min}$ is the
minimum separation at which pairs can be reliably separated (nominally $2''$);
$\theta_{max}$ corresponds to a physical separation, $r_p$, chosen so that:
({\it i}) physical pairs are doomed to merge in $< 10^9$yr (on the basis of both
empirical studies and conventional dynamics); and ({\it ii}) most pairs in the
sample are physically associated. These conditions are satisfied by $r_p\approx 30$ kpc, which corresponds to $\approx 10''$\ at $z = 0.3$, assumed to be the mean
redshift of galaxies in our sample; e.g., the Supernova 1995K was found
at $z = 0.478$ (Leibundgut et al. 1995).

We have used selection criteria based on those of local samples (Ka\-ra\-chen\-tsev 1972; Hickson
1982; Prandoni et al. 1994) in order to define our isolation criterion and to avoid unrelated
galaxies, i.e., optical pairs (Sulentic 1992).  We defined the radius, R$_{G}$,
which is the radius of the smallest circle containing the centers of the group
members (for pairs, R$_{G}$ is half of the pairs separation). We selected  only
pairs and groups with no neighbors within a distance R$_{N}$ to the center of
the group, so that R$_{N}$/R$_{G}$ $\ge$ 3. 

Our group algorithm selects pairs and groups of galaxies
within $19<m_{R}<22$ and $2''<\theta<6''$. However, because of our
completeness limit (see \S3.1) we can only identify galaxies that are
brighter than $m_R=22$. Therefore, groups and pairs which have
neighbors fainter than this limit were still considered as isolated.
Compactness constraint like the one defined by Prandoni et al. were
not used in our group selection since this constraint is less
stringent at faint magnitudes (see Fig. 6 in Prandoni et al. 1994). 
The main difference between our criteria and local samples criteria 
resides in the fact that our membership criterion considers all 
galaxies within the range of magnitude $19<m_{R}<22$ instead of 
considering all galaxies up to three magnitudes fainter than the 
brightest member.

All pairs and groups selected were inspected by eye and classified according to their components intensities and isolation.
We centered each pair and group in a 14$''$ rectangular region  and used a surface plot centered
on the objects to measure their intensities, $I$ (see \S6 for classification).
75 pairs and 1 group were identified as spurious detections and
removed from the catalog.
We have also excluded 0.3942 deg$^{2}$ due to fields overlapping. 
Only galaxies within $19<m_{R}<22$ were considered and, 
therefore, galaxies outside this limit but which fell 
inside the isolation circle were not considered as 
members and the group was still selected. When we classified our
pairs/groups we looked for these objects (see column 10 of
Tables 1 and 2). Pairs/groups classified as {\it 4} are pairs/groups with a faint object very close or with a bright object on the border of a 14$''$
box; pairs/groups with an object within a 14$''$ box were
classified as {\it 5}. Approximately half of the quartets belongs to these classes.

\section{Positions}

The following procedure is used to transform positions, ($x,y$), on the 
CCD images to equatorial coordinates ($\alpha,\delta $) for the equinox 
2000.0. A number of faint stars that are visible in our fields were chosen 
from the Digitized Sky Survey images. An average of 20 stars were 
matched to the CCD positions and were used to find the coefficients of 
second-order polynomials of the following form:

\begin{eqnarray}
\alpha -\alpha_o=\sum_{j=1}^m\sum_{i=1}^ja_{ij}(x-x_o)^{j-i}
(y-y_o)^{i-1} ~,\\
\delta -\delta_o=\sum_{j=1}^m\sum_{i=1}^jb_{ij}(x-x_o)^{j-i}
(y-y_o)^{i-1} ~.
\end{eqnarray}

\noindent where $(x-x_o),~(y-y_o)$ are the CCD centroid coordinates
 from a reference position, and $(\alpha-\alpha_o),~(\delta-\delta_o)$ are
equatorial coordinates relative to the reference coordinates.
A total of 12 independent coefficients defines the
transformation from machine ($x,y$) units to equatorial, ($\alpha,\delta$).  
The accuracy of the fit (as judged by the root mean square residuals
of the fit) was better than 0.5$''$. 

\section{Description of the Catalogs}

In Tables 1 and 2 we present the catalogs of faint pairs and groups of galaxies.
The catalog contains information on all of the objects that are outside the
``excised areas'' discussed in \S2. Five parameters per object are
stored.  The column entries are described below.

\noindent
{\it Columns (1) to (6).} 
Right ascension ($\alpha$) and declination ($\delta$), epoch 2000.

\noindent
{\it Column (7) }
The ``total'' $m_R$ magnitude as defined in \S2, calibrated as described in \S3.
  
\noindent
{\it Column (8) }
Group Radius, R$_{G}$, in arcsecs as defined in \S4. For
pairs, R$_{G}$ means half of the pairs separation.

\noindent
{\it Column (9) }
Isolation parameter, R$_{N}$/R$_{G}$, as defined in \S4. All pairs and 
groups were selected such that R$_{N}$/R$_{G}>3$.  

\noindent
{\it Column (10) }
Pairs and Groups Classification: 1- members with similar intensity 
(I$_{min}\approx$I$_{max}$) and
with no other object within a 14$''$ box; 2- members with I$_{min}$ $> 0.5$ I$_{max}$ and with no other object within a 14$''$ box; 3- members with I$_{min}$ 
$\le$ 0.5 I$_{max}$ and with no other object within a 14$''$ box; 4- pairs or groups with a faint object very close or with a bright object on the border of a 14$''$ box; 5- pairs or groups with an object within a 14$''$ box; 6- pairs with a faint member which could be an HII region of brighter galaxy. I$_{min}$ and I$_{max}$ correspond to members with minimum and maximum intensity; a 14$''$ box was centered in each pair or group (see \S4).

\section{Discussion}

We have found 1751 isolated pairs of galaxies and 30 groups of 
galaxies within $ 19<m_{R}<22$ and $2''<\theta<6''$ out of 73,988 
galaxies in a total area of 2.2333~deg$^{2}$. 
We have performed a simulation in order to compare the number of 
pairs and groups identified by our algorithm with the number predicted by 
galaxies randomly distributed on the sky. Our simulation makes a random 
sample with all the galaxies identified within the 2.616 deg$^{2}$ area,which corresponds to the total area (2.8532 deg$^{2}$) subtracted of the masked area (0.2372 deg$^{2}$). The same pair and group selection criteria described above was applied to the random sample. 

The pair radius, $R_G$ (half the pair separation), and the nearest neighbor
index, $R_N/R_G$, histograms are presented in Figures  \ref{fig:hist-rad-102}
and \ref{fig:hist-rnrg-102}. The dotted line in both histograms
represents detections on the random catalogs and the data used are for
the 2.616 deg$^{2}$ area. Our results show clearly
an increase in pairs and groups of galaxies in comparison to the randomly generated catalog. For instance, at $3''< \theta < 4''$ there are 2
to 3 times more pairs as expected from the random catalog, which in turn, is
an excess of close pairs over what would be expected from an extrapolation of
$\omega$($\theta)$ at larger $\theta$ (Carlberg et al. 1994; Infante
\& Pritchet 1995). Future papers will discuss the interpretation of
these results in order to examine the rise in the merger rate with
redshift (Infante et al. 1996) and a second catalog covering a different
area in the sky has been selected (de Mello et al. 1996). The authors
hope that this work and the fact that these objects are equatorial and accessible from both hemispheres will motivate an effort to obtain redshifts for many objects in these catalogs.

\placefigure{\ref{fig:hist-rad-102}}

\placefigure{\ref{fig:hist-rnrg-102}}

\placetable{tab1}
\placetable{tab2}
 
\acknowledgments
We are grateful to the High-Z Supernovae Search Group for making their images available, to the anonymous referee for helpful comments, to C. Mendes de Oliveira for interesting suggestions, and to 
Gladys Vieira for the careful work in obtaining the coordinates. We
acknowledge the Digitized Sky Survey produced at the Space Telescope
Science Institute under U. S. Government grant NAG W-2166. DFM thanks
FAPESP and CNPq for the fellowships and the hospitality of CTIO and P. Universidad Cat\'olica. 
LI thanks Fondecyt Chile for support through `Proyecto 1930570'.

\clearpage

\figcaption [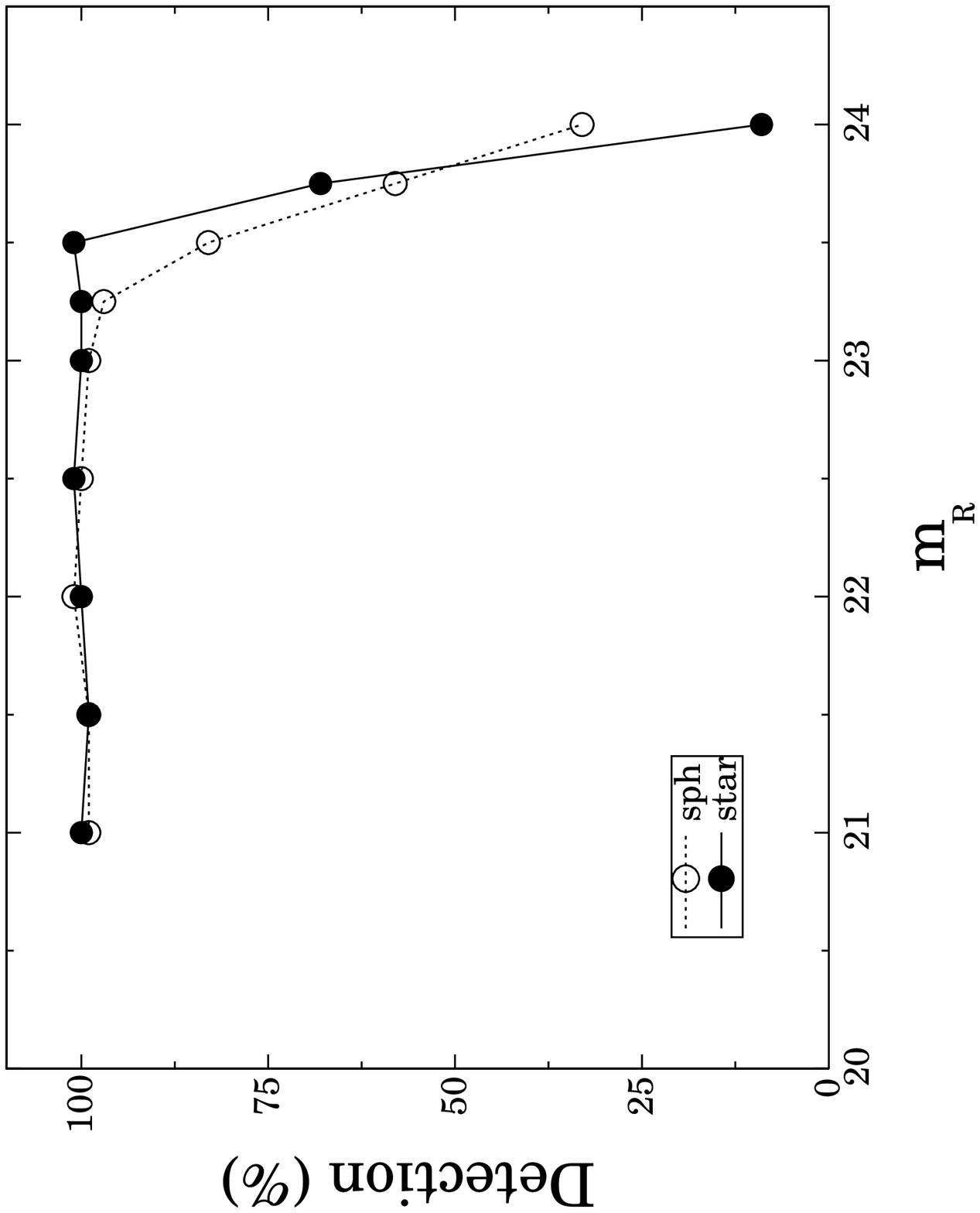]{Percentage of objects (resolved and unresolved)
detected as a function of magnitude m$_{R}$. Filled circles are stars and open
circles are spheroidal galaxies. \label{fig:comp-mag}} 

\figcaption [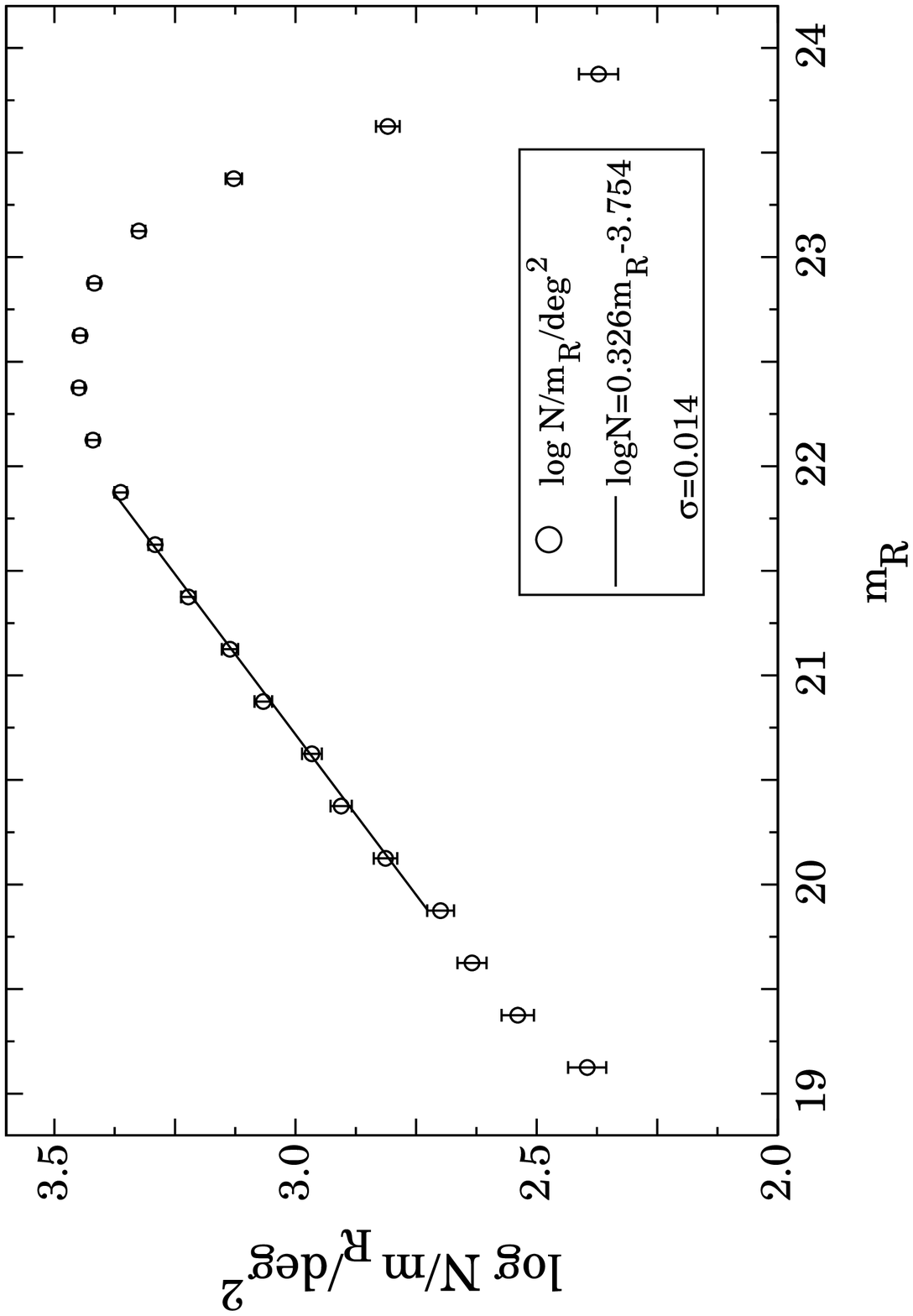]{Galaxy number counts as a function of $m_R$ magnitude.
The poisson error bars are smaller than the symbols in most points. The
turn over in the number counts is due to incompleteness.
\label{fig:counts}}

\figcaption [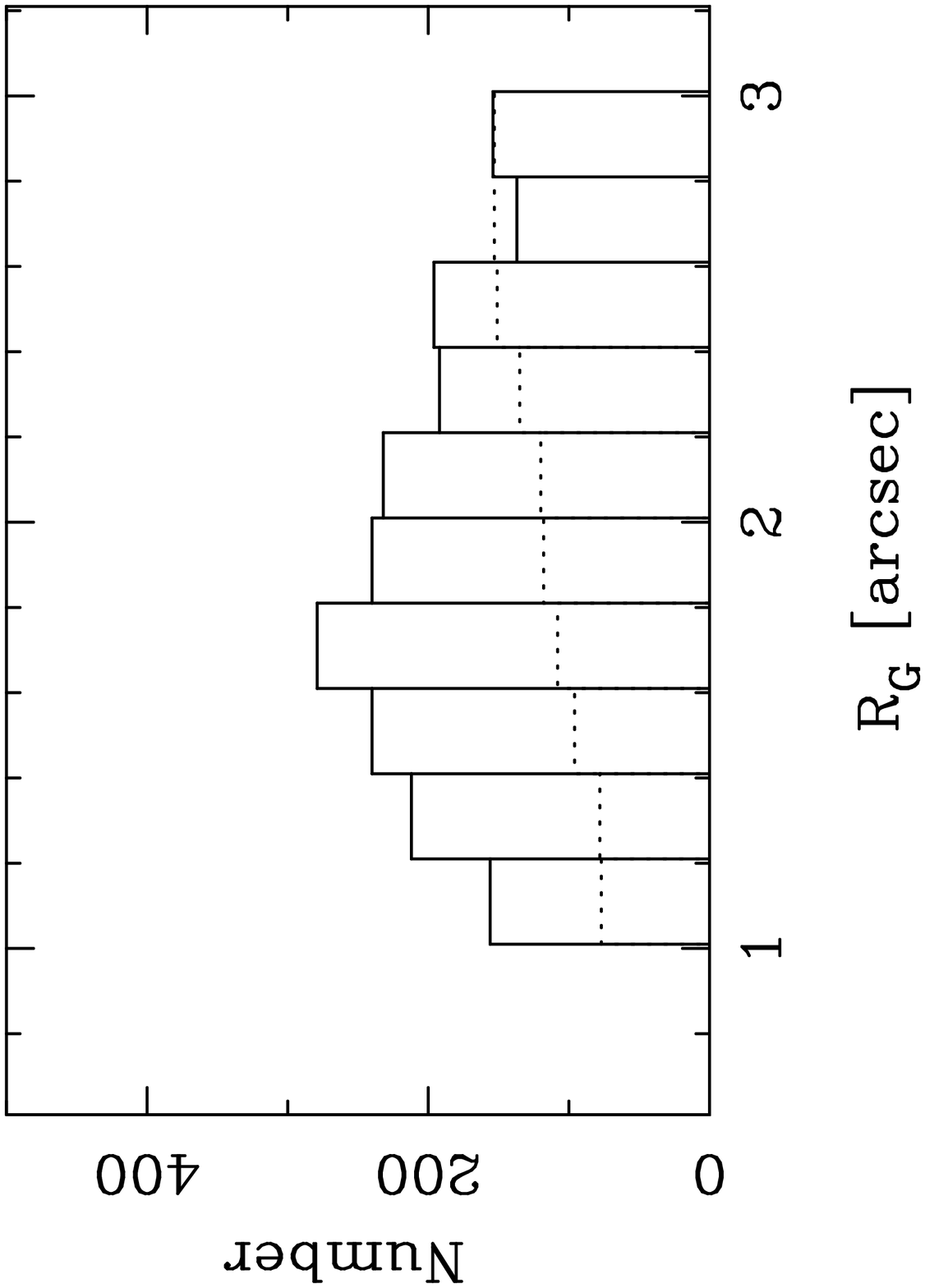]{Histogram of the group radius, $R_G$, for
pairs. The survey data are shown as a solid line and the random catalog as a
dotted line. (See the text for details.) Note that for pairs $R_G$ corresponds
to half the pair separation. \label{fig:hist-rad-102}}

\figcaption [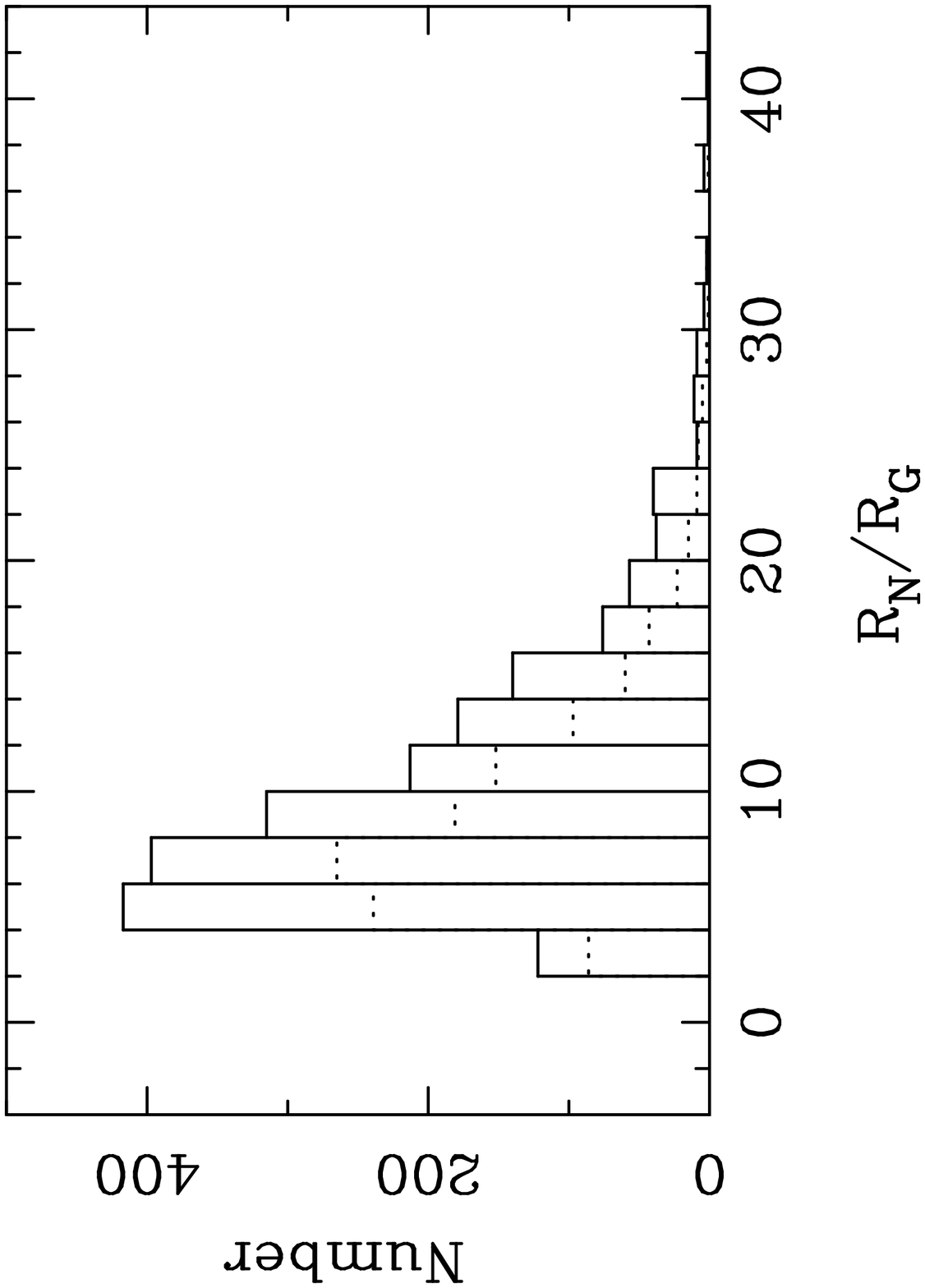]{Histogram of the isolation parameter,
$R_N/R_G$, for pairs. The survey data are shown as a solid line and the random
catalog as a dashed line. (See the text for details.) Pairs and groups with
$R_N/R_G~<~3$ are excluded from the catalog. \label{fig:hist-rnrg-102}}

\clearpage

\plotone{comp-mag.ps}

\plotone{counts.ps}

\plotone{hist-rad-102.ps}

\plotone{hist-rnrg-102.ps}


\clearpage
\begin{thebibliography}{}
\bibitem[]{} Butcher, H., \& Oemler, A. 1984, \apj, 285, 426
\bibitem[]{} Burkey, J. M., Keel, W. C., Windhorst, R. A., \& Franklin, B. E. 1994, ApJ, 429, L13
\bibitem[]{} Carlberg, R. G., Pritchet, C. J., \& Infante, L. 1994, ApJ, 435, 540
\bibitem[]{} de Mello, D. F., Infante, \& L., Menanteau, F. 1996, in preparation
\bibitem[]{} Ellis, R. S. 1995, in Stellar Populations, IAU Symposium 164, eds P. C. van der Kruit \& G. Gilmore (Kluwer), p. 291
\bibitem[]{} Hickson, P. 1982, ApJ, 255, 382
\bibitem[]{} Infante, L. 1987, A\&A, 183, 177
\bibitem[]{} Infante, L., \& Pritchet, C. J. 1995, ApJ, 439, 565
\bibitem[]{} Infante, L., Pritchet, C. J., \& Quintana, H. 1986, AJ,
91, 21
\bibitem[]{} Infante, L., de Mello, D. F., \& Menanteau,
F. 1996, ApJ, submitted
\bibitem[]{} Karachentsev, I. D. 1972, Comm. Spec. Ap. Obs. 7, 1
\bibitem[]{} Koo, D. C., \& Kron, R. G. 1992, \araa, 30, 613
\bibitem[]{} Kron, R. 1980, ApJS, 43, 305
\bibitem[]{} Leibundgut, B., et al. 1995, The Messenger, 81, 19
\bibitem[]{} Maia, M., Pastoriza, M. G., Bica, E., \& Dottori, H. 1994, ApJS, 93, 425
\bibitem[]{} Prandoni, I., Iovino, A., MacGillivray, H. T. 1994, AJ,
107, 1235
\bibitem[]{} Reduzzi, L., \& Rampazzo, R. 1995, Astrophysical Letters \& Communications 30, 1. Gordon and Breach
\bibitem[]{} Schmidt, B. 1995, private communication
\bibitem[]{} Soares, D. S.L., de Souza, R. E., de Carvalho, R. R., \& Couto da Silva, T. C. 1995, A\&ASS, 110, 371
\bibitem[]{} Sulentic, J. W. 1992, in Morphological and Physical Classification of Galaxies,  eds G. Longo, M. Capaccioli \& G. Busarello (Kluwer), p. 293 
\bibitem[]{} Yee, H. K. C. \& Ellingson, E. 1995, ApJ, 445, 37
\bibitem[]{} Zepf, S. E., \& Koo, D. C. 1989, ApJ, 337, 34
\end{thebibliography}
\end{document}